\magnification \magstep1
\raggedbottom
\openup 2\jot
\voffset6truemm
\headline={\ifnum\pageno=1\hfill\else
\hfill {\it Noncovariant Gauges in Simple Supergravity}
\hfill \fi}
\centerline {\bf NONCOVARIANT GAUGES IN}
\centerline {\bf SIMPLE SUPERGRAVITY}
\vskip 1cm
\centerline {GIAMPIERO ESPOSITO}
\centerline {\it Istituto Nazionale di Fisica Nucleare,
Sezione di Napoli,}
\centerline {\it Mostra d'Oltremare Padiglione 20,
80125 Napoli, Italy;}
\centerline {\it Dipartimento di Scienze Fisiche,}
\centerline {\it Mostra d'Oltremare Padiglione 19,
80125 Napoli, Italy}
\vskip 0.3cm
\centerline {GIUSEPPE POLLIFRONE}
\centerline {\it Dipartimento di Fisica, Universit\`a di
Roma ``La Sapienza"}
\centerline {\it and INFN, Sezione di Roma, Piazzale Aldo
Moro 2, 00185 Roma, Italy}
\vskip 1cm
\noindent
A gauge-averaging functional of the axial type is studied for
simple supergravity at one loop about flat Euclidean four-space
bounded by a three-sphere, or two 
concentric three-spheres. This is
a generalization of recent work on the axial gauge in quantum
supergravity on manifolds with boundary. Ghost modes obey
nonlocal boundary conditions of the spectral type, in that
half of them obey Dirichlet or Neumann conditions at the boundary.
In both cases, they give a vanishing contribution to the
one-loop divergence. The admissibility of noncovariant gauges
at the classical level is also proved.
\vskip 4cm
\leftline {PACS numbers: 04.65.+e, 98.80.Hw}
\vskip 100cm
\leftline {\bf 1. Introduction}
\vskip 0.3cm
\noindent
The recent investigations of boundary conditions and heat-kernel
asymptotics in Euclidean quantum gravity have led to new work on
the application of noncovariant gauges in the quantization of
Einstein's gravity and simple supergravity.$^{1-3}$ The motivations
of such an analysis are as follows.
\vskip 0.3cm
\noindent
(i) To improve the understanding of the one-loop semiclassical
approximation. If quantum theory is viewed as a theory of small
disturbances of the underlying classical theory, one should
be able to compute at least the first quantum corrections in
powers of $\hbar$, if such a perturbative scheme can be of any
help (despite the well known lack of perturbative 
renormalizability). 
\vskip 0.3cm
\noindent
(ii) To obtain a complete picture of admissible boundary
conditions for the quantization of gauge fields and gravitation.
Boundary conditions are here viewed as an essential element of
any quantization scheme, as is suggested by recent progress in
Euclidean quantum gravity.$^{3,4}$ 
\vskip 0.3cm
\noindent
(iii) To ensure self-adjointness of the elliptic operators 
acting on graviton and gravitino perturbations, when a
problem with boundary is studied.$^{1-3}$  
\vskip 0.3cm
\noindent
(iv) To understand whether supergravity theories are at least
one-loop finite in the presence of boundaries.$^{2,3}$

In simple supergravity, which is the object of our investigation,
it is by now well known that one has a choice of local or
nonlocal boundary conditions on gravitino perturbations. The
former may involve complementary projection operators at the
boundary,$^{5}$ or may fix on the initial surface the whole primed
part of tangential components of gravitino perturbations, and
on the final surface the whole unprimed part of tangential 
components of gravitino perturbations.$^{6,7}$ The latter rely
instead on the following scheme.$^{2,3}$ The {\it massless}
Rarita-Schwinger potential subject to gauge conditions and 
linearized supersymmetry constraints is split into a regular part
and a singular part. The regular part may be written as an
infinite sum of modes multiplying harmonics having positive
eigenvalues of the intrinsic three-dimensional Dirac operator
of the boundary. The singular part is instead an infinite sum
of modes multiplying harmonics having negative eigenvalues
of the intrinsic three-dimensional Dirac operator of the
boundary. One thus performs a nonlocal operation, i.e. the
separation of the spectrum of a first-order elliptic operator
into its positive and negative part.$^{2,8}$ This is closely 
related to a positive- and negative-frequency split, which is
suitable both for scattering theory and one-loop quantum 
cosmology.$^{6,7}$ 

These nonlocal boundary conditions of the spectral type for
gravitino perturbations can be written in the form$^{2}$ 
$$
\Bigr[\psi_{i(+)}^{A}\Bigr]_{\partial M}=0 \; ,
\eqno (1.1)
$$
$$
\Bigr[{\widetilde \psi}_{i(+)}^{A'}\Bigr]_{\partial M}=0 \; ,
\eqno (1.2)
$$
where the label $(+)$ denotes the part of the perturbation 
potential corresponding to the regular part of the 
underlying classical theory. In Eqs. (1.1) and (1.2) one deals
with the tangential components of the perturbation potential.
Their expression in terms of spinor and tetrad fields is
$$
\psi_{i}^{A}=\Gamma_{\; \; \; \; \; B}^{C'A}
\; e_{\; \; C'i}^{B} \; ,
\eqno (1.3)
$$
$$
{\widetilde \psi}_{i}^{A'}=\gamma_{\; \; \; \; \; B'}^{CA'}
\; e_{\; \; \; Ci}^{B'} \; ,
\eqno (1.4)
$$
where $\Gamma$ and $\gamma$ are the purely two-spinor part
of Rarita-Schwinger potentials, while $e_{\; \; C'i}^{B}$
is the spatial component of the two-spinor version of the
tetrad. The infinitesimal gauge transformations for
Rarita-Schwinger potentials are$^{2}$ 
$$
{\widehat \Gamma}_{\; \; \; BC}^{A'}
\equiv \Gamma_{\; \; \; BC}^{A'}
+\nabla_{\; \; \; B}^{A'} \; \nu_{C} \; ,
\eqno (1.5)
$$
$$
{\widehat \gamma}_{\; \; B'C'}^{A} \equiv
\gamma_{\; \; B'C'}^{A}+\nabla_{\; \; B'}^{A}
\; \mu_{C'} \; .
\eqno (1.6)
$$
To ensure invariance of the boundary conditions (1.1) and
(1.2) under such gauge transformations, one begins by 
considering the conditions$^{2}$ 
$$
\Bigr[\nabla_{\; \; \; B}^{A'} \; \nu_{C_{(+)}}
\Bigr]_{\partial M}=0 \; ,
\eqno (1.7a)
$$
$$
\Bigr[\nabla_{\; \; B'}^{A} \; \mu_{C'_{(+)}}
\Bigr]_{\partial M}=0 \; .
\eqno (1.8a)
$$
In the (one-loop) quantum theory, the spinor fields $\nu_{C}$
and $\mu_{C'}$ should be regarded as the ghost fields, and
the boundary conditions (1.7a) and (1.8a) do not lead, by
themselves, to a well-posed problem for ghost perturbations,
since each of them leads to 8 conditions at the boundary. To
overcome this problem, the work in Ref. 2 considered a particular
subsector, obtained by contracting Eq. (1.7a) with the Euclidean
normal ${_{e}n_{\; \; A'}^{B}}$, and Eq. (1.8a) with the Euclidean
normal ${_{e}n_{A}^{\; \; B'}}$. As shown in Ref. 2, this
prescription leads to a peculiar set of spectral boundary
conditions on ghost modes, 
in that half of them (i.e. the regular ones) 
obey Neumann conditions at the boundary. 

However, when gauge-averaging functionals of the axial type
are chosen, another possibility consists
in contracting Eqs. (1.7a) and (1.8a) with that particular
linear combination of Euclidean normals which then sets to
zero at the boundary the action of the ghost operators on
$\nu^{C}$ and $\mu^{C'}$, respectively. For example, if the
axial-type gauge functionals are chosen as in Ref. 2:
$$
\Phi^{A}(\Gamma) \equiv {_{e}n_{CC'}} \; \Gamma^{ACC'} \; ,
\eqno (1.9)
$$
$$
{\widetilde \Phi}^{A'}(\gamma) \equiv 
{_{e}n_{CC'}} \; \gamma^{A'C'C} \; ,
\eqno (1.10)
$$
one finds the ghost operators$^{2}$ 
$$
{\cal D}_{C}^{\; \; A} \equiv 
{_{e}n_{CC'}} \; \nabla^{AC'} \; ,
\eqno (1.11)
$$
$$
{\cal F}_{C'}^{\; \; \; A'} \equiv
{_{e}n_{CC'}} \; \nabla^{CA'} \; ,
\eqno (1.12)
$$
which, of course, act linearly on $\nu^{C}$ and 
$\mu^{C'}$, respectively. On the other hand, contraction of
Eq. (1.7a) with ${_{e}n_{\; \; A'}^{C}}$ leads to
$$
\Bigr[{\cal D}_{CB} \; \nu_{(+)}^{C}\Bigr]_{\partial M}=0 \; .
\eqno (1.13)
$$
An analogous procedure for Eq. (1.8a) leads to
$$
\Bigr[{\cal F}_{C'B'} \; \mu_{(+)}^{C'}
\Bigr]_{\partial M}=0 \; .
\eqno (1.14)
$$
At a deeper level, Eq. (1.13) is obtained by requiring
that the boundary conditions
$$
\Bigr[\Phi^{A}(\Gamma)\Bigr]_{\partial M}=0
\eqno (1.15a)
$$
should be invariant under the infinitesimal gauge
transformations (1.5):
$$
\Bigr[\Phi^{A}({\widehat \Gamma})\Bigr]_{\partial M}=0 \; .
\eqno (1.15b)
$$
Similarly, Eq. (1.14) is obtained if one requires that the
boundary conditions
$$
\Bigr[{\widetilde \Phi}^{A'}(\gamma)\Bigr]_{\partial M}=0
\eqno (1.16a)
$$
should be preserved by the infinitesimal gauge transformations
(1.6):
$$
\Bigr[{\widetilde \Phi}^{A'}({\widehat \gamma})
\Bigr]_{\partial M}=0
\; \; \; \; .
\eqno (1.16b)$$
Moreover, by virtue of the linear action 
of the ghost operators, Eqs. (1.13)
and (1.14) are equivalent to spectral boundary conditions where
all regular ghost modes are set to zero on the final surface, if
the initial three-surface shrinks to a point (as it happens in the
Hartle-Hawking approach to quantum cosmology${}^{9}$).

In our paper, we consider a more general class of 
gauge-averaging functionals of the axial type. They are
defined as
$$
\chi^{A}(\Gamma) \equiv {_{e}n_{CC'}} \;
\Gamma^{(AC)C'} \; ,
\eqno (1.17)
$$
$$
{\widetilde \chi}^{A'}(\gamma) \equiv
{_{e}n_{CC'}} \; \gamma^{(A'C')C} \; ,
\eqno (1.18)
$$
where round brackets denote symmetrization over spinor indices
of the same kind. In Ref. 2 it was pointed out that the corresponding
one-loop properties were still unknown and should have been
analyzed. It has been therefore our aim to 
study a Faddeev-Popov path-integral
representation for the one-loop wave function of the universe 
which involves a Gaussian average over the gauge functionals 
(1.17) and (1.18), in the limit of small three-geometry.$^{10}$ 
The corresponding background can be taken to be a portion of
flat Euclidean four-space bounded by a 
three-sphere of radius $a$,$^{3,10}$ with 
radial coordinate $\tau \in [0,a]$.

Section 2 derives ghost operators and boundary conditions when
the gauge-averaging functionals (1.17) and (1.18) are chosen. 
One-loop properties are studied
in Sec. 3, and the admissibility of our noncovariant
gauges at the classical level is proved in Sec. 4. 
Concluding remarks are presented in Sec. 5.
\vskip 0.3cm
\leftline {\bf 2. Ghost Operators and Eigenvalue Equations}
\vskip 0.3cm
\noindent
In the Faddeev-Popov path integral for simple
supergravity in the axial gauge, 
one has to perform a gaussian average involving 
the left-hand sides
of (1.17) and (1.18), jointly with 
their corresponding ghost terms.
 
In the case of massless gravitino perturbations, 
as we said in the introduction, 
the ghost operators are obtained 
by studying the behaviour of (1.17) and (1.18) 
under the infinitesimal 
gauge transformations (1.5) and (1.6), respectively. 
Thus, the ghost operators acting on the spinor fields $\nu^{C}$ 
and $\mu^{C'}$ read 
$$
{{\cal P}^A}_{C} \equiv 
\Bigr({_{e}n_{CC'}}{\nabla^{AC'}}  -
{{\varepsilon}^{A}}_{C}
\; {_{e}n_{FC'}{{\nabla}^{FC'}}}\Bigr)\; ,
\eqno (2.1)
$$
and 
$$
{{\cal R}^{A'}}_{C'} \equiv 
\Bigr({_{e}n_{CC'}}{{\nabla}^{CA'}}  -
{{\varepsilon}^{A'}}_{C'} 
\; {_{e}n_{CF'}}{{\nabla}^{CF'}}\Bigr)\;.
\eqno (2.2)
$$
One can write the eigenfunction expansions for the spinor fields
$\nu^C$ and $\mu^{C'}$ as$^{2}$
$$
\nu^{C} = \sum_{\lambda}{\cal B}^{1}_{\lambda}
\;\nu^{C}_{(\lambda)}\; ,
\eqno (2.3)
$$
$$
\mu^{C'} = \sum_{\tilde\lambda}{\cal B}^{2}_{\tilde\lambda}
\;\mu^{C'}_{({\tilde\lambda})}\; ,
\eqno (2.4)
$$
and then analyze the following eigenvalue equations:
$$
{{\cal P}^A}_{C} \; \nu^{C}_{(\lambda)}=
\lambda \nu^{A}_{(\lambda)}\; ,
\eqno (2.5)
$$
$$
{{\cal R}^{A'}}_{C'} \; \mu^{C'}_{({\tilde\lambda})}=
{\tilde\lambda} \mu^{A'}_{({\tilde\lambda})}\; .
\eqno (2.6)
$$
It is well known that for massless gravitinos, 
expressed in the form 
(1.3) and (1.4) and subject to the gauge transformations
(1.5) and (1.6), the background is forced to be Ricci-flat.$^{11}$ 
This happens since the spinor fields $\nu_{C}$ and 
$\mu_{C'}$ have to be freely specifiable inside 
the background four-manifold. Considering furthermore 
a local description of Rarita-Schwinger potentials 
occurring in (1.3) and (1.4) in terms of a second set 
of potentials, one finds that the background 
with boundary is further 
restricted to be totally flat.$^{2,12}$

We here study a flat Euclidean background bounded by 
a three-sphere. Thus, bearing in mind that$^{2}$
$$
{_{e}n_{A A'}}\nabla^{A A'} \nu^{C}_{(\lambda)}
= -{\partial \over \partial \tau}\nu^{C}_{(\lambda)}\; ,
\eqno (2.7)
$$
the left-hand sides of
Eqs. (2.5) and (2.6) take the form (cf. Ref. 2)
$$
{{\cal P}^A}_{C} \; \nu^{C}_{(\lambda)}=
-\Bigr({3}
{\partial \over \partial \tau}\nu^{A}_{(\lambda)}
+2\;{_{e}n_{C'}}^{(C}e^{A)C'i} \; {^{(4)}\nabla_{i}}
\nu_{C({\lambda})}\Bigr)\; ,
\eqno (2.8)
$$
and 
$$
{{\cal R}^{A'}}_{C'} \; \mu^{C'}_{({\tilde\lambda})}=
-\Bigr({3}
{\partial \over \partial \tau}\mu^{A'}_{(\tilde\lambda)}
+2\;{_{e}n_{C}}^{(C'}e^{A')Ci} \; {^{(4)}\nabla_{i}}
\mu_{C'({\tilde\lambda})}\Bigr)\; ,
\eqno (2.9)
$$
respectively. Further details about the notation 
can be found in Ref. 2.

One now has to impose boundary 
conditions for the solutions of the eigenvalue equations 
(2.5) and (2.6). A suitable set of boundary conditions  
consist in the vanishing of the axial-type
gauge-averaging functionals (1.17) and (1.18) at the boundary. 
This choice can be viewed as the generalization of magnetic 
boundary conditions of Euclidean Maxwell
theory.$^3$ More precisely, for fields of spin 
$1,{3\over 2}$ and $2$ one can always set to zero at the
boundary the gauge-averaging functional (either covariant,
e.g. Lorentz, harmonic or de Donder, or noncovariant, e.g.
axial or Coulomb). This is part of a set of mixed boundary
conditions for the quantum theory.$^{1-4}$ The requirement
that such boundary conditions should be preserved 
under the infinitesimal gauge transformations 
(1.5) and (1.6) leads to 
$$
\left[{{\cal P}^A}_{C} \; \nu^{C}_{(+)}\right]_{\partial M} =0\; ,
\eqno (2.10)
$$  
and
$$
\left[{{\cal R}^{A'}}_{C'}
\; \mu^{C'}_{(+)}\right]_{\partial M} =0\; .
\eqno (2.11)
$$ 
\vskip 0.3cm
\leftline {\bf 3. One-loop Analysis in the Axial Gauge}
\vskip 0.3cm 
\noindent
In this section we study the one-loop properties of our ghost fields 
in flat Euclidean background bounded by a three-sphere. 
For this purpose, we expand the ghost fields in harmonics 
on a family of three-spheres centred on the origin$^{2}$ 
$$
\nu^{A}=\sum_{n=0}^{\infty}\sum_{p,q=1}^{(n+1)(n+2)}
\alpha_{n}^{pq}\Bigr[m_{np}(\tau)\rho^{nqA}
+{\widetilde r}_{np}(\tau){\overline \sigma}^{nqA}\Bigr] \; ,
\eqno (3.1)
$$
$$
\mu^{A'}=\sum_{n=0}^{\infty}\sum_{p,q=1}^{(n+1)(n+2)}
\alpha_{n}^{pq}\Bigr[{\widetilde m}_{np}(\tau)
{\overline \rho}^{nqA'}+r_{np}(\tau)\sigma^{nqA'}\Bigr] \; ,
\eqno (3.2)
$$
where the $\alpha_{n}$ are block-diagonal matrices with blocks
$\pmatrix {1&1 \cr 1&-1 \cr}$, 
and the harmonics obey the
eigenvalue equations described, for example, in Ref. 2.
Hence one finds that 
the eigenvalue equations (2.5) and (2.6) for the 
ghost modes occurring in the 
expansions (3.1) and (3.2) read 
$$
-\Bigr(3{d\over d\tau}+{n\over \tau}\Bigr)m_{np}
=\lambda_{n} \; m_{np} \; ,
\eqno (3.3)
$$
$$
-\Bigr(3{d\over d\tau}-{(n+3)\over \tau}\Bigr)
{\widetilde r}_{np}=\lambda_{n} \; {\widetilde r}_{np} \; ,
\eqno (3.4)
$$
$$
-\Bigr(3{d\over d\tau}-{n\over \tau}\Bigr)
{\widetilde m}_{np}={\widetilde \lambda}_{n} \;
{\widetilde m}_{np} \; ,
\eqno (3.5)
$$
$$
-\Bigr(3{d\over d\tau}+{(n+3)\over \tau}\Bigr)r_{np}
={\widetilde \lambda}_{n} \; r_{np} \; .
\eqno (3.6)
$$
The solutions of Eqs. (3.3)--(3.6) are (cf. Eqs. 
(2.21)--(2.24) in Ref. 2)
$$
m_{np}(\tau,\lambda_{n})=\tau^{-{n\over 3}} \; 
e^{-{\lambda_{n}\over 3}\tau}
\; m_{np}^{0} \; ,
\eqno (3.7)
$$
$$
{\widetilde r}_{np}(\tau,\lambda_{n})=\tau^{{(n+3)\over 3}}
\; e^{-{\lambda_{n}\over 3}\tau} \; {\widetilde r}_{np}^{0} \; ,
\eqno (3.8)
$$
$$
{\widetilde m}_{np}(\tau,{\widetilde \lambda}_{n})
=\tau^{n\over 3} \; e^{-{{\widetilde \lambda}_{n}\over 3} \tau}
\; {\widetilde m}_{np}^{0} \; ,
\eqno (3.9)
$$
$$
r_{np}(\tau,{\widetilde \lambda}_{n})
=\tau^{-{(n+3)\over 3}} \; e^{-{{\widetilde \lambda}_{n}\over 3} \tau}
\; r_{np}^{0} \; ,
\eqno (3.10)
$$
where $m_{np}^{0},{\widetilde r}_{np}^{0},
{\widetilde m}_{np}^{0}$ and $r_{np}^{0}$ denote some
multiplicative constants which are determined 
by the boundary values.
Remarkably, regularity at the origin is guaranteed by the vanishing 
of the $m_{np}$ and $r_{np}$ modes, for all $\tau \in [0,a]$, 
where $a$ is the radius
of the three-sphere boundary. Moreover, one has to 
impose the boundary conditions 
(2.10) and (2.11) on the remaining (regular) 
set of ghost modes. Since the ghost operators act linearly on ghost
modes (see (2.5) and (2.6)), 
such spectral boundary conditions set to zero everywhere 
the ${\widetilde m}_{np}$ and 
${\widetilde r}_{np}$ modes. Hence the whole set of ghost
modes are forced to vanish, with this version of the
boundary conditions in the axial gauge. 

By contrast, the spectral boundary conditions studied in 
Ref. 2 and mentioned in the introduction:
$$
\left[{_{e}n_{\; \; A'}^{B}} \; \nabla_{\; \; \; B}^{A'}
\; \nu_{C_{(+)}} \right]_{\partial M}=0 \; ,
\eqno (1.7b)
$$
$$
\left[{_{e}n_{A}^{\; \; B'}} \; \nabla_{\; \; B'}^{A}
\; \mu_{{C'}_{(+)}} \right]_{\partial M}=0 \; ,
\eqno (1.8b)
$$
lead to Neumann boundary conditions 
at the boundary, i.e.
$$
\left[{d \over d\tau}{\widetilde r}_{np}
\right]_{\tau=a}=0 
\eqno (3.11)
$$
and
$$
\left[{d \over d\tau}{\widetilde m}_{np}
\right]_{\tau=a}=0 \; .
\eqno (3.12)
$$
One thus gets, from (3.8) and (3.9), the same 
discrete spectra for the regular modes found in Ref. 2,   
although the axial gauge-averaging functionals
(1.9) and (1.10) differ from (1.17) and (1.18)
considered in our paper. Such discrete spectra are
$$
\lambda_{n}={(n+3)\over a} \; \; \; \; \forall n \geq 0 \; ,
\eqno (3.13)
$$
$$
{\widetilde \lambda}_{n}={n\over a} \; \; \; \; \forall n \geq 0 \; .
\eqno (3.14)
$$
Setting for convenience $a=1$, one can define the following 
$\zeta$-function for a first-order, elliptic and
positive-definite operator $\cal A$  
(cf. (2.8)) with spectrum (3.13):
$$
\zeta_{\cal A}(s) \equiv \sum_{n=0}^{\infty}
(n+1)(n+2)(n+3)^{-s} \; .
\eqno (3.15)
$$
Moreover, a first-order, elliptic and nonnegative operator $\cal B$
exists (cf. (2.9)), with spectrum (3.14) and {\it finite-dimensional} 
null-space, and its 
$\zeta$-function can be defined as
$$
\zeta_{\cal B}(s) \equiv 2 + \sum_{n=1}^{\infty}
(n+1)(n+2)n^{-s} \; ,
\eqno (3.16)
$$
where the dimension of the null-space has been included in the
definition of the $\zeta$-function, following Ref. 2.
One then finds (here, $\zeta_{H}$ and $\zeta_{R}$ are the
Hurwitz and Riemann $\zeta$-functions, respectively,
defined in the appendix of Ref. 2) 
$$
\zeta_{\cal A}(0)=\zeta_{H}(-2,3)-3\zeta_{H}(-1,3)
+2 \zeta_{H}(0,3)=-{3\over 4} \; ,
\eqno (3.17)
$$
$$
\zeta_{\cal B}(0)=2+\zeta_{R}(-2)+3\zeta_{R}(-1)
+2 \zeta_{R}(0)={3\over 4} \; .
\eqno (3.18)
$$
Hence the ghost gravitino contribution to the one-loop divergence,
in both versions of the axial gauge, vanishes in our flat background
bounded by a three-sphere:
$$
\zeta_{\rm ghost}(0)=\zeta_{\cal A}(0)+\zeta_{\cal B}(0)=0 \; .
\eqno (3.19)
$$
Note that no Nielsen-Kallosh ghost fields$^{13,14}$ occur in
our calculation, since, as explained in Ref. 2, the axial gauge
has already the effect to reduce the linearized gravitino
potential to its two physical degrees of freedom, corresponding
to helicities ${3\over 2}$ and $-{3\over 2}$.

So far, motivated by quantum cosmology, we have studied backgrounds 
where the initial three-surface shrinks to zero. However, in quantum 
field theory one deals with a path-integral representation 
of transition amplitudes with suitable data on {\it two} 
boundary three-surfaces. Hence we now analyze 
ghost modes of simple supergravity on a flat Euclidean 
four-manifold bounded by two
concentric three-spheres. The form of ghost modes is again 
(3.7)--(3.10), and the spectral boundary conditions 
resulting from (1.7b) and (1.8b) are
$$
\left[{d \over d\tau}{m}_{np}\right]_{\tau=\tau_{-}}=0 \; ,
\eqno (3.20)
$$
$$
\left[{d \over d\tau}{r}_{np}\right]_{\tau=\tau_{-}}=0 \; ,
\eqno (3.21)
$$
$$
\left[{d \over d\tau}{\widetilde m}_{np}
\right]_{\tau=\tau_{+}}=0 \; ,
\eqno (3.22)
$$
$$
\left[{d \over d\tau}{\widetilde r}_{np}
\right]_{\tau=\tau_{+}}=0 \; ,
\eqno (3.23)
$$
where $\tau_-$ and $\tau_+$ are the three-sphere radii  
($\tau_{+} > \tau_-$). Similarly to the analysis of Ref. 2,
such boundary conditions lead to incompatible solutions for 
the eigenvalues $\lambda_n$ and ${\widetilde\lambda}_n$. In particular, 
(3.20) and (3.23) imply that $\lambda_{n}=-{n\over \tau_{-}} \leq 0$
and $\lambda_{n}={(n+3)\over \tau_{+}} > 0$, while (3.21)
and (3.22) lead to 
${\tilde \lambda}_{n}=-{(n+3)\over \tau_{-}}<0$ and
${\tilde \lambda}_{n}={n\over \tau_{+}} \geq 0$. Hence 
no nontrivial ghost modes exist. This remains true
if one uses instead the spectral boundary conditions
(2.10) and (2.11) when $\tau=\tau_{-}$ and $\tau=\tau_{+}$,
by virtue of (2.5), (2.6) and (3.7)--(3.10).

Following Refs. 1,2, the one-loop divergence for simple
supergravity in the axial gauge reduces therefore to the
contribution resulting from three-dimensional
transverse-traceless perturbations for gravitons and
gravitinos:
$$
\zeta(0)=-{278\over 45}+{289\over 360}=-{43\over 8} \; ,
\eqno (3.24)
$$
when only one bounding three-sphere exists, and
$$
\zeta(0)=-5 \; ,
\eqno (3.25)
$$
when two concentric three-sphere boundaries occur.
\vskip 5cm
\leftline {\bf 4. Classical Admissibility of Axial Gauges}
\vskip 0.3cm
\noindent
The axial gauge-averaging functionals play an important role 
in the path-integral approach to quantum gravity, since 
they lead to self-adjoint operators on metric 
and gravitino perturbations, and are part of the quantization
programme in noncovariant gauges.$^{15-20}$ This makes it
interesting to investigate the classical counterpart of the
axial gauge-averaging functionals (1.17) and (1.18). Hence
we impose, in the underlying classical theory, the axial 
gauge, i.e.
$$
{_{e}n_{CC'}} \; \Gamma^{(AC)C'}=0 \; ,
\eqno (4.1)
$$
$$
{_{e}n_{CC'}} \; \gamma^{(A'C')C} \; .
\eqno (4.2)
$$
The preservation of the gauge conditions (4.1) and (4.2)
under the infinitesimal gauge transformations (1.5)
and (1.6) leads
to the following differential equations for the spinor 
fields $\nu^A$ and $\mu^{A'}$:
$$
{_{e}n_{CC'}} \; \nabla^{C'(A}\nu^{C)}=0 \; ,
\eqno (4.3)
$$
$$
{_{e}n_{CC'}} \; \nabla^{C(A'}\mu^{C')}=0 \; .
\eqno (4.4)
$$
Hence the spinor fields $\nu^A$ and $\mu^{A'}$ are no longer 
freely specifiable, but they have to satisfy Eqs. (4.3) and (4.4).
Interestingly, a {\it sufficient} condition for the validity
of such equations is obtained if $\nu^{A}$ and $\mu^{A'}$
are solutions of the twistor equation.$^{12}$ 
 
The restriction on gauge fields resulting from the preservation
of gauge conditions should not be surprising, and is indeed 
a familiar property in field theory.
For example, it is well known that (Euclidean) Maxwell theory 
is invariant under 
the following infinitesimal gauge transformations:
$$
{}^{f}\!A_{b} = A_{b} + \nabla_{b} f \; ,
\eqno (4.5)
$$
where $A_b$ is the electromagnetic potential and 
the function $f$ is freely specifiable. 
However, after imposing a gauge condition, the function $f$ 
has to obey a differential equation, instead of being 
freely specifiable. On choosing the axial gauge:
$$
n^{b}A_{b}=0 \; ,
\eqno (4.6) 
$$
it is possible to ensure that 
the gauge-transformed potential 
(4.5) does actually obey the gauge (4.6), provided that 
$n^{b}\nabla_{b}f=0$. In the case of the Lorentz gauge:
$$
\nabla^{b}A_{b}=0 \; ,
\eqno (4.7)
$$
the preservation of Eq. (4.7) under (4.5) 
forces the function $f$ to obey the four-dimensional 
Laplace equation, if the background is Riemannian.
Another choice is the Coulomb gauge, i.e. 
$$
{}^{(3)}\nabla^{i}A_{i}=0 \; ,
\eqno (4.8)
$$
where ${}^{(3)}\nabla^{i}$ is the three-dimensional 
tangential covariant derivative 
with respect to the Levi-Civita connection of the induced 
metric on the boundary. On requiring that the tangential components  
${ }^{f}\! A_{i}$ of the gauge-transformed potential should 
satisfy Eq. (4.8), one finds that the function $f$ has to be a 
harmonic function on the boundary. Thus, provided that  
a gauge condition is imposed, 
and such a gauge condition also holds for the potential
${}^f\! A_{b}$, the function $f$ is no longer freely specifiable, just
as in the case of the axial gauge for the Rarita-Schwinger potential.

Now, it is convenient to study a mode-by-mode 
form of Eqs. (4.3) and (4.4), by using 
the expansions (3.1) and (3.2). Hence one finds 
(cf. Eqs. (3.3)--(3.6))
$$
\Bigr(3{d\over d\tau}+{n\over \tau}\Bigr)m_{np}
=0 \; ,
\eqno (4.9)
$$
$$
\Bigr(3{d\over d\tau}-{(n+3)\over \tau}\Bigr)
{\widetilde r}_{np}= 0 \; ,
\eqno (4.10)
$$
$$
\Bigr(3{d\over d\tau}-{n\over \tau}\Bigr)
{\widetilde m}_{np}=0 \; ,
\eqno (4.11)
$$
$$
\Bigr(3{d\over d\tau}+{(n+3)\over \tau}\Bigr)r_{np}
=0 \; .
\eqno (4.12)
$$
The solutions of Eqs. (4.9)--(4.12) can be written as
$$
m_{np}(\tau)=\tau^{-{n\over 3}} \; 
m_{np}^{0} \; ,
\eqno (4.13)
$$
$$
{\widetilde r}_{np}(\tau)=\tau^{{(n+3)\over 3}}
\; {\widetilde r}_{np}^{0} \; ,
\eqno (4.14)
$$
$$
{\widetilde m}_{np}(\tau)
=\tau^{n\over 3} 
\; {\widetilde m}_{np}^{0} \; ,
\eqno (4.15)
$$
$$
r_{np}(\tau)
=\tau^{-{(n+3)\over 3}} 
\; r_{np}^{0} \; .
\eqno (4.16)
$$
To ensure regularity at the origin 
it is necessary to set to zero everywhere 
the modes $m_{np}$ and $r_{np}$. Hence we have proved that the 
axial gauge is admissible at the classical level, provided 
that the irregular modes occurring in the 
expansions in harmonics  
(3.1) and (3.2) vanish for all $\tau \in [0,a]$. 
As far as the regular modes are concerned, one can point out 
that Eqs. (4.3) and (4.4) should hold everywhere, and hence, 
in particular, at the boundary. This, however, does not fix 
${\widetilde r}_{np}$ and ${\widetilde m}_{np}$. 
\noindent
\vskip 0.3cm
\leftline {\bf 5. Concluding Remarks}
\vskip 0.3cm
\noindent
The contributions of our paper are as follows.
\vskip 0.3cm
\noindent
(i) It has been shown that any
choice of axial gauge-averaging functional
is compatible with having the standard form of spectral
boundary conditions on ghost modes, i.e. when half of them
are set to zero at the boundary. The work in Ref. 2 only 
considered the Neumann option for such modes.
\vskip 0.3cm
\noindent
(ii) The most general form of the ghost operators resulting
from a gauge-averaging functional of the axial type has been
obtained in Eqs. (2.1) and (2.2). The corresponding ghost
contribution to the one-loop divergence vanishes. Thus,
all gauges of the axial type are equivalent in the quantum
theory. 
\vskip 0.3cm
\noindent
(iii) The admissibility of the axial gauge 
for problems with boundary has been proved
at the classical level. One starts from a configuration where
the axial gauge is fulfilled. One then performs a gauge 
transformation on the Rarita-Schwinger potentials. This
restricts the class of spinor fields $\nu_{C}$ and $\mu_{C'}$
occurring in Eqs. (1.5) and (1.6). They are no longer freely
specifiable. They have instead to solve Eqs. (4.3)
and (4.4). The mode-by-mode form of such equations has been
given and solved in the local coordinates appropriate for the
case when a portion of flat Euclidean four-space is bounded
by a three-sphere.

Our investigations are part of a more general programme, 
devoted to the study of Euclidean quantum gravity and
quantum supergravity in covariant and noncovariant gauges
(cf. Refs. 15--20),
when boundary effects are included.$^{1-3}$ As shown in Refs.
3--5, there is increasing evidence that this programme
is going to shed new light on the various approaches to the
quantization of gauge theories, and on the fertile interplay
between spectral geometry and quantum field theory.

As far as supergravity theories are concerned, a further line
of development lies in the application to supersymmetric
quantum cosmology. So far, the main emphasis has been on the
application of Hamiltonian methods in such a branch of
modern quantum cosmology (see Refs. 6,21 and references
therein). It now remains to be seen whether the techniques 
applied in Refs. 1--3 and in our paper can improve the
current understanding of the quantum state of the universe
within a supersymmetric framework.
\vskip 0.3cm
\leftline {\bf Acknowledgments}
\vskip 0.3cm
\noindent
We are indebted to Ivan Avramidi and Alexander Kamenshchik
for scientific collaboration on topics related to the
one studied in our paper.
\vskip 0.3cm
\leftline {\bf References}
\vskip 0.3cm
\noindent
\item {1.}
I. G. Avramidi, G. Esposito and A. Yu. Kamenshchik, 
{\it Class. Quantum Grav.} {\bf 13}, 2361 (1996).
\item {2.}
G. Esposito and A. Yu. Kamenshchik, {\it Phys. Rev.}
{\bf D54}, 3869 (1996).
\item {3.}
G. Esposito, A. Yu. Kamenshchik and G. Pollifrone, 
{\it Euclidean Quantum Gravity on Manifolds with Boundary},
Fundamental Theories of Physics 
(Kluwer, Dordrecht, 1997).
\item {4.}
I. G. Moss and P. J. Silva, {\it Phys. Rev.} {\bf D55}, 1072 
(1996).
\item {5.}
H. C. Luckock, {\it J. Math. Phys.} {\bf 32}, 
1755 (1991).
\item {6.}
P. D. D'Eath, {\it Supersymmetric Quantum Cosmology}
(Cambridge University Press, Cambridge, 1996).
\item {7.}
G. Esposito, {\it Phys. Lett.} {\bf B389}, 510 (1996).
\item {8.}
M. F. Atiyah, V. K. Patodi and I. M. Singer, 
{\it Math. Proc. Camb. Phil. Soc.} {\bf 77}, 43 (1975).
\item {9.}
J. B. Hartle and S. W. Hawking, {\it Phys. Rev.}
{\bf D28}, 2960 (1983).
\item {10.}
K. Schleich, {\it Phys. Rev.} {\bf D32}, 1889 (1985).
\item {11.}
S. Deser and B. Zumino, {\it Phys. Lett.} 
{\bf B29}, 335 (1976).
\item {12.} 
G. Esposito, G. Gionti, A. Yu. Kamenshchik, I. V. Mishakov
and G. Pollifrone, {\it Int. J. Mod. Phys.} {\bf D4},
735 (1995).
\item {13.}
N. K. Nielsen {\it Nucl. Phys.} {\bf B140}, 499 (1978).
\item {14.} 
R. E. Kallosh {\it Nucl. Phys.} {\bf B141}, 141 (1978).
\item {15.}
T. Matsuki, {\it Phys. Rev.} {\bf D19}, 2879 (1979).
\item {16.}
T. Matsuki, {\it Phys. Rev.} {\bf D21}, 899 (1980).
\item {17.}
D. M. Capper and G. Leibbrandt, {\it Phys. Rev.}
{\bf D25}, 1009 (1982).
\item {18.}
D. M. Capper and G. Leibbrandt, {\it Phys. Rev.}
{\bf D25}, 2211 (1982).
\item {19.}
T. Matsuki, {\it Phys. Rev.} {\bf D32}, 3164 (1985).
\item {20.}
G. Leibbrandt, {\it Rev. Mod. Phys.} {\bf 59},
1067 (1987).
\item {21.}
P. Moniz, {\it Int. J. Mod. Phys.} {\bf A11},
4321 (1996).

\bye